\documentclass[conference]{IEEEtran}
\IEEEoverridecommandlockouts

\usepackage{cite}
\usepackage{amsmath,amssymb,amsfonts}
\usepackage{algorithmic}
\usepackage{graphicx}
\usepackage{textcomp}
\usepackage[dvipsnames]{xcolor}
\usepackage{listings}
\lstdefinestyle{commands}{
  basicstyle=\footnotesize\ttfamily,
  frame=none
}
\usepackage{url}
\usepackage{tikz}
\newcommand{\redcircle}[1]{%
  \tikz[baseline=(char.base)]{
    \node[circle, fill=red, text=white, font=\bfseries, inner sep=1pt] (char) {#1};
  }%
}

\definecolor{ballblue}{rgb}{0.13, 0.55, 0.95}

\lstdefinelanguage{json}{
    morestring=[b]",
    morecomment=[l]{//},
    morekeywords={true,false,null, CPCON},
    sensitive=true,
    linewidth=\linewidth
}

\def\BibTeX{{\rm B\kern-.05em{\sc i\kern-.025em b}\kern-.08em
    T\kern-.1667em\lower.7ex\hbox{E}\kern-.125emX}}
\begin{document}

\title{Towards Centralized Orchestration of Cyber Protection Condition (CPCON)}

\author{\IEEEauthorblockN{1\textsuperscript{st} Mark Timmons}
\IEEEauthorblockA{\textit{Department of Computer Science} \\
\textit{Naval Postgraduate School}\\
Monterey, CA \\
mark.timmons@nps.edu}
\and
\IEEEauthorblockN{2\textsuperscript{nd} Daniel Lukaszewski}
\IEEEauthorblockA{\textit{Department of Computer Science} \\
\textit{Naval Postgraduate School}\\
Monterey, CA \\
dflukasz@nps.edu}
\and
\IEEEauthorblockN{3\textsuperscript{rd} Geoffrey Xie}
\IEEEauthorblockA{\textit{Department of Computer Science} \\
\textit{Naval Postgraduate School}\\
Monterey, CA \\
xie@nps.edu}
\and
\IEEEauthorblockN{4\textsuperscript{th} Thomas Mayo}
\IEEEauthorblockA{\textit{Defensive Cyberspace Operations Warfare Tactics Instructor} \\
\textit{Naval Information Forces}\\
Suffolk, VA \\
thomas.j.mayo18.mil@us.navy.mil}
\and
\IEEEauthorblockN{5\textsuperscript{th} Donald McCanless}
\IEEEauthorblockA{\textit{Defensive Cyberspace Operations Warfare Tactics Instructor} \\
\textit{Naval Postgraduate School}\\
Monterey, CA \\
donald.mccanless@nps.edu}
}

\maketitle

\begin{abstract}
The United States Cyber Command (USCYBERCOM) Cyber Protection Condition (CPCON) framework mandates graduated security postures across Department of Defense (DoD) networks, but current implementation remains largely manual, inconsistent, and error-prone. This paper presents a prototype system for centralized orchestration of CPCON directives, enabling automated policy enforcement and real-time threat response across heterogeneous network environments. Building on prior work in host-based intrusion response, our system leverages a policy-driven orchestrator to standardize security actions, isolate compromised subnets, and verify enforcement status. We validate the system through emulated attack scenarios, demonstrating improved speed, accuracy, and verifiability in CPCON transitions with human-in-the-loop oversight.
\end{abstract}

\begin{IEEEkeywords}
CPCON, Security Orchestration, Network Automation, Intrusion Detection and Response
\end{IEEEkeywords}

\section{Introduction} \label{Intro}

Protecting critical network infrastructure is essential for any organization. The 2023 United States Cyber Command (USCYBERCOM) Command Challenge Problem Set identifies numerous technology areas in need of development through partnerships with industry \cite{uscybercom_ccp_3rd}. Cybersecurity threat detection and mitigation is one such area, where anomaly detection, endpoint hardening, and proactive defense are critical capabilities for countering cyberspace threats.

As an important step toward assured cyber defense, USCYBERCOM has established the Cyber Protection Conditions (CPCON) framework to standardize network protection priorities during cyberspace events \cite{uscybercom_cpcon}. 
CPCON Level 5 is the least restrictive, permitting all network functions. In contrast, CPCON Level 1 is the most restrictive, prioritizing the preservation of critical services while blocking or disabling non-critical ones. This graduated protection model enables network operators to isolate non-essential services, thereby reducing the attack surface and safeguarding mission-critical functions.  

However, we observe that the current practice of CPCON in the fleet involves numerous \emph{ad hoc} manual processes and as such, suffers from the following drawbacks:
\begin{itemize}

\item {\bf Inconsistent implementation}: Network operators with different platforms will implement CPCON measures differently as they draw from their own experience, creating gaps in overall security posture.

\item {\bf Delayed response}: Manual processes will slow down the implementation of some crucial security changes, increasing vulnerability windows.

\item {\bf Lack of compliance verification and reporting}: \emph{Ad hoc} manual configuration is error-prone. Additionally, confirming actual compliance becomes difficult, relying on self-reporting rather than objective metrics.

\item {\bf Training burden}: Personnel require extensive training to understand and implement nuanced CPCON changes across various systems.
\end{itemize}


\begin{figure} [t]
    \centering
    \includegraphics[scale=0.28, trim=10 3 3 5, clip]{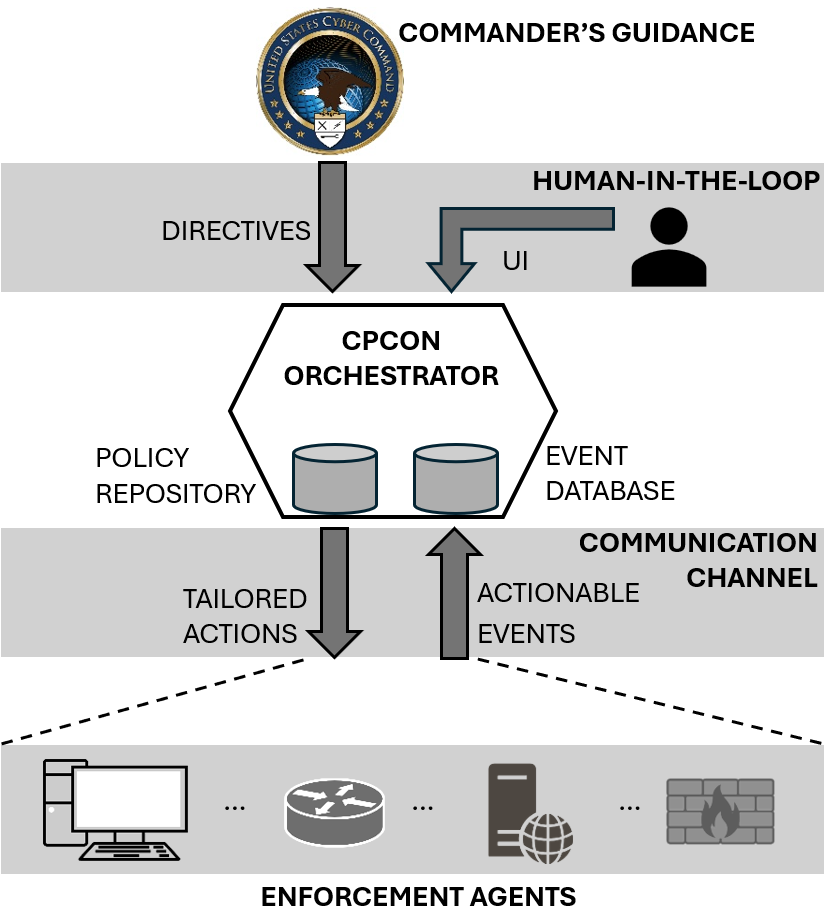}
    \caption{Illustration of proposed centralized orchestration system.}
    \label{fig:CPCON_architecture}
    \vspace{-2mm}
\end{figure}

Meanwhile, our previous work \cite{secsoft2025} demonstrated an iterative capability for deploying software modules in real time, upon observation of malicious events, to mitigate cyber attacks. This functionality was enabled using a centralized orchestrator to process alert information from intrusion detection system (IDS) agents embedded in various devices and, if necessary, dispatch tailored security remedy modules to the devices. In this paper, we extend that concept by enhancing the orchestrator to facilitate standardizing and automating the implementation of CPCON. 

Fig.~\ref{fig:CPCON_architecture} illustrates a high level architectural view of the proposed system. At the bottom layer, CPCON enforcement agents are installed in various devices of the network. A centralized CPCON orchestrator in the middle layer processes alerts from the enforcement agents in real time and dispatches tailored responses in the form of new security modules as instructed by prescribed policy.  Finally, recognizing that \mbox{CPCON} actions should build in flexibility to support command specific discretions, we include mechanisms such as a user interface (UI) at the top layer, for not only taking CPCON directives from USCYBERCOM, but also supporting operator interactions with the system. 

Guided by the architecture, we have created a system prototype,
which includes sample CPCON policy rules specified in JSON format, security modules for controlling access at the granularity of subnets, users, and ports, and a simple operator UI. Furthermore, we have evaluated the prototype in a simulated scenario involving a network with multiple security postures.  




The rest of the paper is organized as follows. In Section~\ref{Related Work}, we discuss related works and how our work differs or builds upon the concepts. Section~\ref{sec:concept} presents our prototype implementation while Sections~\ref{Design of Experiments} and \ref{Results} describe our experimental design and results, respectively.  We discuss potential system extensions and conclude the paper in Sections~\ref{Discussion} and \ref{Conclusion}. 
\section{Related Work}\label{Related Work}

The general concept of security orchestration has been explored to unify detection, mitigation, and response, reducing operator burden and enabling rapid, policy-driven defense \cite{SecurityOrchestration}. Cross-domain event correlation and closed-loop architectures further enhance network resilience by enabling real-time policy updates that preempt threat propagation \cite{AutoSecurity}.
As cloud and mobile technologies expand the threat surface, zero trust architectures are emerging to replace traditional perimeter-based models. Recent work highlights the value of dynamic, host-informed policy orchestration to address gaps in distributed network defense \cite{redefiningZTA,secsoft2025}. Enforcement strategies have also shifted toward user/host-specific controls, such as multi-policy configurations~\cite{IoT_policy_enforcement} and adaptive trust profiles~\cite{jin2021zero}.

At the device level, automated collection of host-based alerts is key to enforcing network-wide security policies and supporting CPCON-level decisions. In~\cite{SnortSVM}, the authors enhance host intrusion detection by combining Snort with support vector machine-based classification to improve alert fidelity. Koyya \cite{Splunk} complements this with a scalable syslog-to-Splunk architecture that enables centralized analysis of distributed alerts. These approaches support real-time correlation and actionable insights for operators. We build on this by using Ansible for automated policy enforcement due to its scalability, remote execution capabilities, and future integration with natural language tooling~\cite{AnsibleLightspeed}.


To support dynamic orchestration in response to evolving threats, AI-driven models can generate actionable signatures to inform orchestration frameworks by enabling automated, targeted responses \cite{automationZTA}. Systems for continuous monitoring and event correlation, when enhanced with deep learning techniques, offer improved anomaly detection—particularly in identifying insider threats \cite{automationZTA}. Our work lays the foundation for integrating such AI-guided orchestration into CPCON policy enforcement.
\section{System Prototyping}\label{sec:concept}


This section presents a system prototype that we have built for centralized orchestration of CPCON. First, we introduce sample declarative rules for standardizing the orchestrator's responses to CPCON directives and cyber events reported from devices. Then, we describe how we developed the orchestrator prototype 
and a sample of CPCON enforcement modules by extending our previous work.



\subsection{Standardized CPCON Rules}\label{sec:rules}

Standardized rules are necessary to remove ambiguity when either providing direction to the CPCON orchestrator or when network devices send event data to the orchestrator for correlation and response.  To enable these operations we designed a structured event and response security policy with a collection of rules in a declarative language format~\cite{Ponder_policy} \cite{DeclarativePolicy}.

CPCON directives, with a rule format shown in Fig.~\ref{fig:directive_format}, are used to enforce a change in the CPCON posture within the network.  The ``CPCON\_level" argument sets the target level for the network. The ``threat" argument provides amplifying information to the specific threat for the given level in order to address specific threats (e.g., phishing).  Finally, the ``action" argument is a list of actions for the orchestrator to automatically enable (e.g., isolate a subnet).

\begin{figure}[htb]
    \centering
    \begin{lstlisting}[basicstyle=\scriptsize, linewidth=\linewidth]
"Directive":
    "CPCON_level": <level 1-5>,
    "threat": <category>,
    "action": {<action_1, host_id(s)>, 
               <action_2, host_id(s)>,...,
               <action_n, host_id(s)>}
    \end{lstlisting}
    \caption{CPCON Directive message format highlighting syntax. 
    }
    \label{fig:directive_format}
\end{figure}


Event statistics generated from host-level alerts can be ingested by the orchestrator to drive adaptive, network-wide security responses. For example, at lower CPCON levels, a single host reporting access to a suspicious website may not prompt action. However, if the CPCON level increases, the same website may be added to a shared event database and dynamically pushed to perimeter gateways as part of a blacklist policy. In more advanced cases, integrated threat intelligence feeds may confirm that the website is linked to command-and-control activity—such as botnet coordination—prompting the orchestrator to issue higher CPCON level recommendations and enforce broader containment or mitigation actions.

The format for events, Fig.~\ref{fig:event_message_format}, is used for network hosts providing information to the CPCON orchestrator.  The ``alert" argument provides the specific alert observed by the host (e.g., unauthorized connection).  The ``CPCON\_level" argument is included to provide host-context for the alert and the ``action" argument provides automatic actions taken by the host (e.g., block connection). 

\begin{figure}[htb]
    \centering
    \begin{lstlisting}[basicstyle=\scriptsize]
"Event": 
   "alert": <host_id, event_type>, 
   "CPCON_level": <level 1-5>,
   "action": {<response_1>,<response_2>,..., <response_n>}
    \end{lstlisting}
    \caption{CPCON Event message format highlighting syntax. 
    }
    \vspace{-2mm}
    \label{fig:event_message_format}
\end{figure}

\subsection{CPCON Orchestrator and Enforcement Modules}\label{system_design}

\begin{figure}[htb]
    \centering
    \includegraphics[width=0.65\linewidth]{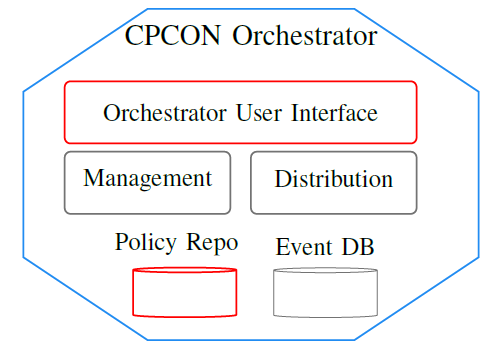}
    \caption[Layer 4.5 Orchestrator.]{Adapted from \cite{secsoft2025, lukaszewski2023software}. Modified Layer 4.5 orchestrator consists of new policy components to facilitate CPCON directives (red).  The existing components (grey) are leveraged to deploy policy throughout the network.}
    \label{fig:orchestrator}
\end{figure}

The Layer4.5 orchestrator developed in our previous work\cite{lukaszewski2023software} included multiple functions to establish channels with network devices and support the deployment of customization modules. In a follow-up effort~\cite{secsoft2025}, we adapted the orchestrator to deploy security-focused modules aimed at enhancing intrusion detection and response. In this work, we leverage those existing functions to deploy CPCON policies across the network, as illustrated in Fig.~\ref{fig:orchestrator}.

\begin{figure}[htb]
    \vspace*{2mm}
    \centering
    \includegraphics[scale=0.49, trim=0 2 2 5, clip]{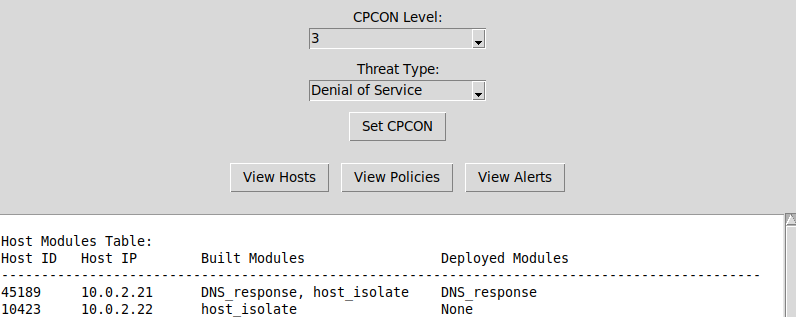}
    \caption{Orchestrator UI Prototype}
    \label{fig:Orechstrator_UI}
    \vspace{-4mm}
\end{figure}

The key enhancements to the orchestrator are the UI, Fig.~\ref{fig:Orechstrator_UI}, and the policy repository for enforcing CPCON directives.  Critical to individual network owners is the ability to view real time alerts, ordered directives, and policy requirement validation.  The user interface provides a means for the human-in-the-loop to i) view current policy enforcement, ii) monitor events, and iii) automate policy enforcement verification.


The enforcement modules used in this work are implemented as kernel object files, built and deployed from the orchestrator~\cite{lukaszewski2023software}. These modules enable fine-grained monitoring of socket-level traffic to detect specific threat signatures and anomalous behavior. Upon identifying suspicious patterns, the modules generate alerts that are forwarded to the orchestrator for correlation and response.

In this paper we leverage the new UI with associated newly developed security modules to implement a series of CPCON changes as we progress through a scenario designed to showcase automation when responding to cyber threats.

\section{Design of Experiments}\label{Design of Experiments}
This section presents a set of experiments that we have developed for evaluating the orchestration system prototype. The presentation focuses the testbed architecture, experimentation objectives, scenarios and threat vectors modeled, and the performance metrics to be analyzed.

\subsection{Testbed Design}

\begin{figure}[htb]
    \centering
    \includegraphics[scale=0.45, trim=10 5 10 5, clip]{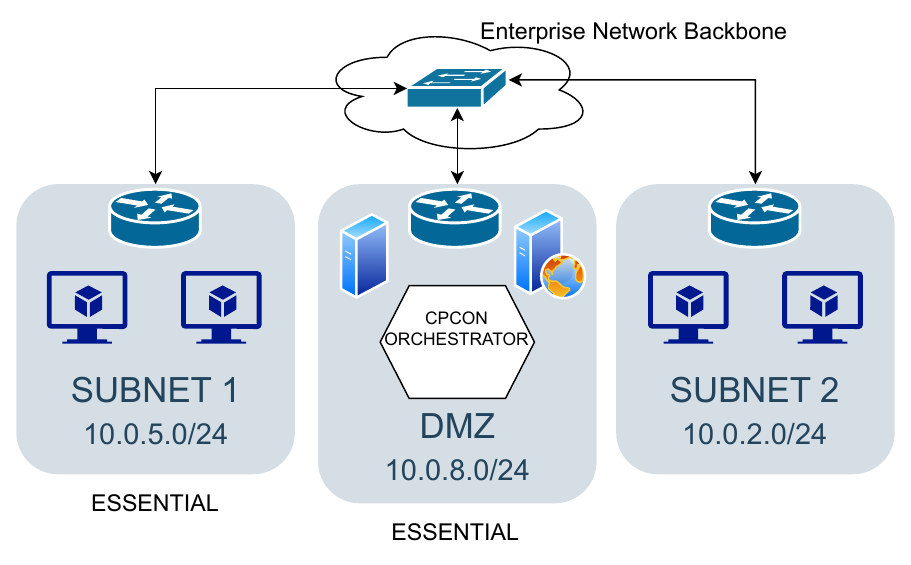}
    \caption{Network topology used for experimentation. Subnet~1 and the DMZ are designated as ``essential", while Subnet~2 is ``non-essential". 
    }
    \label{fig:Network}
\end{figure}

Fig.~\ref{fig:Network} illustrates the network topology used to model an enterprise environment with segmented subnets to simulate varying levels of mission criticality. Subnet~1 and the DMZ are designated as \textit{essential}, while Subnet~2 represents \textit{non-essential} infrastructure. This classification supports CPCON-level enforcement by enabling selective isolation of less critical assets under elevated threat conditions.

Subnet~1 and Subnet~2 host two generic virtualized hosts, while the DMZ includes the CPCON orchestrator, a web server, and a utility server offering common network services. All inter-subnet routing is performed via dedicated Ubuntu-based virtual routers configured with \texttt{iptables} firewall rules. The orchestrator leverages Ansible to remotely configure the routers in accordance with CPCON directives.

The testbed is physically hosted on three Windows~11 laptops, each running Ubuntu~20.04 virtual machines (VMs) with kernel version 5.13 and Layer~4.5~\cite{lukaszewski2023software} support enabled. Each system is equipped with an Intel Core~i7 processor, 16~GB of RAM, and SSD storage. The laptops are interconnected via a TL-SG108E managed 1~Gbps Ethernet switch.

\begin{figure*}
    \centering
    \includegraphics[width=0.9\linewidth]{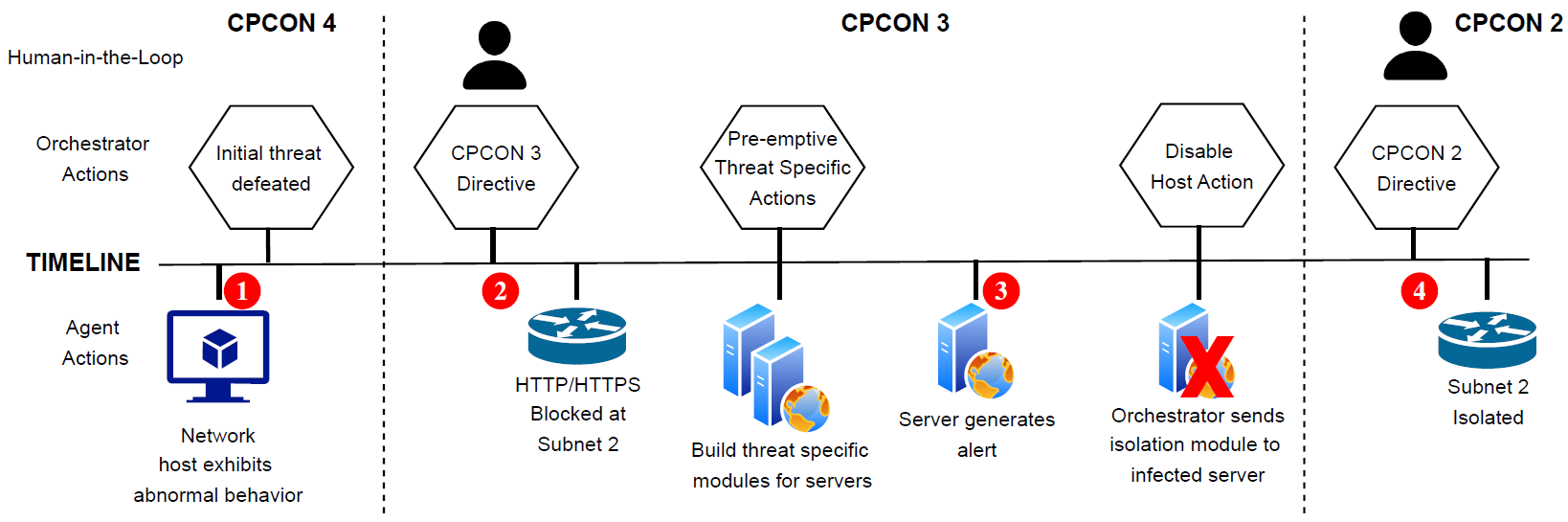}
    \caption{Timeline of scenario events. Orchestrator actions are above and host behavior is displayed below the timeline. Key events: DoS attack from host \protect\redcircle{1}, CPCON level 3 ordered \protect\redcircle{2}, threat specific alert generated and sent to CPCON orchestrator \protect\redcircle{3}, and CPCON level 2 ordered \protect\redcircle{4}.}
    \label{fig:timeline}
\end{figure*}

\subsection{Adversarial Model and Experimental Objectives} \label{Model and Objective}
To evaluate centralized orchestration in an intrusion prevention system (IPS)-like role, two adversarial scenarios based on the MITRE ATT\&CK framework are simulated. These scenarios reflect realistic cyber threats requiring automated defensive responses to protect critical network assets.

The first scenario involves a DoS attack originating from the non-essential subnet, characterized by a sudden spike in DNS queries from a single host—suggestive of malicious activity \cite{mitre_t1498}. 
In the second scenario, following an elevated CPCON level, a DMZ-based web server initiates an outbound connection using a non-standard source port, which may indicate lateral movement or data exfiltration~\cite{MITRE_T1571, intro_networks}. 

These attack scenarios allow validating the orchestrator’s ability to (1) automate threat response, (2) correlate alerts to support CPCON escalation, and (3) enforce network-wide defenses, thereby demonstrating its role in adaptive, layered cybersecurity posture management.

\subsection{Test Scenarios and Configurations}\label{Test Configs}

Our experimentation follows the timeline provided in Fig.\ref{fig:timeline} and begins at CPCON Level 4 in response to an anticipated threat, despite the absence of observed malicious activity within the network. This intentional deviation from standard CPCON escalation procedures allows us to showcase system capabilities under controlled experimental conditions. 

The first scenario models a DNS-based DoS attack launched by a compromised host in Subnet~2. A Python script emulates malicious behavior by issuing high-rate DNS queries to the DMZ utility server, while a local response module monitors for query rate anomalies. Upon detecting anomalous DNS traffic via host-based alert, the orchestrator initiates tailored actions as follows:

\begin{enumerate}
    \item Deploys a DNS DoS mitigation module to alerting host.
    \item The module confirms the attack, rate-limits the process, and notifies the orchestrator.
    \item The event is logged and CPCON escalation is recommended.
\end{enumerate}

Shortly after the DoS attack and in response to receiving elevated threat intelligence, a human-in-the-loop operator raises CPCON to level 3 using the orchestrator UI. Upon receiving this CPCON directive (Fig.~\ref{fig:directive_example}), the orchestrator:

\begin{enumerate}
    \item Blocks HTTP/HTTPS traffic to/from Subnet 2 (non-essential).
    \item Builds and deploys monitoring modules for essential servers targeting specific behaviors, such as the use of ephemeral source ports.
    \item Preemptively builds a host isolation module for all managed hosts.
\end{enumerate}

\begin{figure}[htb]
    \centering
    \begin{lstlisting}[basicstyle=\scriptsize]
"Directive":
    "CPCON_level" : <3>,
    "threat" : <web_applications>,
    "action" : {<Block_web_traffic, subnet2>, 
                <Server_monitor, all_servers>,
                <Build_isolate_mod, all_hosts>}
    \end{lstlisting}
    \caption{CPCON directive message transitioning to Level 3.}
    \vspace{-2mm}
    \label{fig:directive_example}
\end{figure}


After receiving the monitoring module, the DMZ web server attempts an outbound connection using an ephemeral port, matching known threat indicators included in the enforcement module.  The web server immediately notifies the orchestrator, which responds by:

\begin{enumerate}
    \item Deploying the isolation module to remove the web server from the network.
    \item Recommending CPCON escalation to Level~2.
\end{enumerate}

Upon CPCON~2 directive issuance (Fig.~\ref{fig:directive_cpcon2}), the orchestrator isolates the non-critical subnet and confirms enforcement via a follow-up scan. 
Operators are notified of successful implementation.

\begin{figure}[htb]
    \centering
    \begin{lstlisting}[basicstyle=\scriptsize]
"Directive":
    "CPCON_level" : <2>,
    "threat" : <web_attacks>,
    "action" : {<isolate, subnet2>}
    \end{lstlisting}
    \caption{CPCON Directive message transitioning to CPCON level 2 for Phase 3 of experimentation scenario.  
    }
    \vspace{-1em}
    \label{fig:directive_cpcon2}
\end{figure}

\subsection{Performance Metrics} \label{Objectives}
For this prototype implementation, the automation of CPCON directive execution serves as the primary performance metric. 
The objective is to employ defensive measures capable of countering malicious network threats in a manner that demonstrates the standardization of automation relative to traditional human response. Additional evaluation metrics include the orchestrator’s ability to self-assess the successful implementation of CPCON directives, thereby preventing false reports of policy enforcement.

\section{Results}\label{Results}

This section presents the experimental results in a narrative format aligned with the attack scenarios described in Section~\ref{Test Configs} and Fig.~\ref{fig:timeline}. 

\begin{figure}[h]
    \centering
    \vspace{2mm}
    \includegraphics[scale=0.48, trim=0 2 2 5, clip]{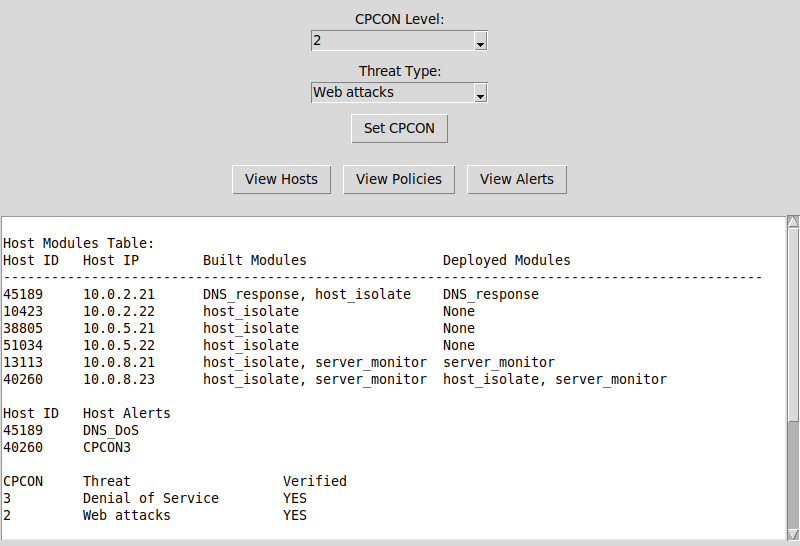}
    \caption{Orchestrator UI showing final state.  Status of each host, detected alerts, and policy implementation and verification are visible.}
    \label{fig:Final_state}
    \vspace{-1.5em}
\end{figure}

Upon detecting an abnormal DNS query rate, the security enforcement module logs a ``DNS\_DoS" alert in the event database. This triggers the CPCON orchestrator to build and deploy a ``DNS\_response" module to the alerting host, as shown in Fig.~\ref{fig:Final_state} for Host ID 45189.  Once deployed, the module enforces DNS query rate limiting, effectively neutralizing the resource exhaustion threat and restoring service stability.


In response to the evolving threat, a directive is issued to elevate to CPCON level 3. A human-in-the-loop operator initiates this transition via the CPCON orchestrator UI. The orchestrator subsequently executes the following Ansible playbook command to enforce the new policy:

\vspace{-0.5em}
\begin{center}
   \begin{lstlisting}[style=commands]
ansible-playbook -i $SUBNET2_router, WEB_Block.yaml
\end{lstlisting} 
\end{center}
\vspace{-0.5em}

\begin{figure}[h]
    \centering
    \begin{tikzpicture}
        \node[anchor=south west,inner sep=0] at (0,0) {\includegraphics[width=0.95\linewidth]{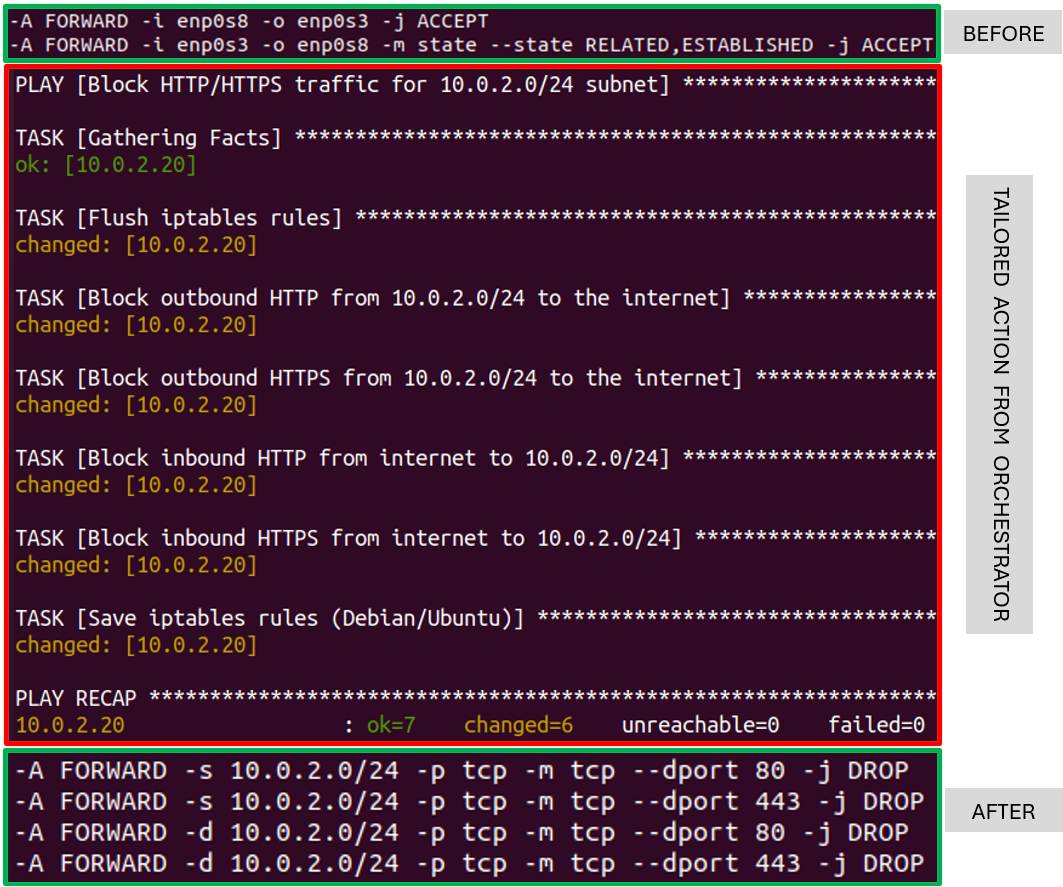}};
        
    \end{tikzpicture}
    \caption{Deployment of Ansible playbook to block HTTP/HTTPS traffic at Subnet 2 router via \texttt{iptables} update. Center red box shows Ansible playbook successfully running, with top and bottom green boxes showing before and after router configurations, respectively.}
    \label{fig:WEB_Block_conbined}
\end{figure}

The output resulting from this operation is shown in Fig.~\ref{fig:WEB_Block_conbined}, confirming the successful reconfiguration of firewall policies to restrict HTTP/HTTPS traffic within the non-essential subnet.

\begin{figure}[h]
    \centering
    \begin{tikzpicture}
        \node[anchor=south west,inner sep=0] at (0,0) {\includegraphics[width=0.95\linewidth]{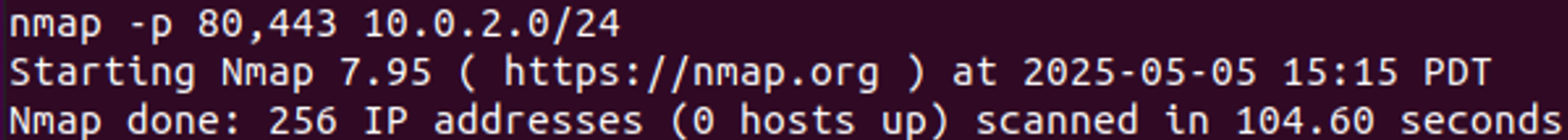}};
        
    \end{tikzpicture}
    \caption{Verification of no hosts are accepting inbound connections on port 80 (HTTP) and 443 (HTTPS).}
    \label{fig:WEB_Block_verify}
    \vspace{-1em}
\end{figure}

Verification was performed using an Nmap scan issued from the CPCON orchestrator to the Subnet~2 router.  The scan confirmed that the mitigation module was successfully deployed and that the expected service restrictions were enforced. As a result, the policy repository was updated to mark the CPCON~3 directive as \textit{verified}, seen in the bottom section of Fig.~\ref{fig:Final_state}. The nmap command performed is as follows:

\begin{lstlisting}[style=commands]
sudo nmap -Pn -p 80,443 $SUBNET2
\end{lstlisting}

Subsequently, a compromised web server attempts to initiate an unauthorized outbound connection. This activity is intercepted and blocked by the response module deployed during the CPCON~3 configuration with a follow-up alert ``CPCON3" sent to the orchestrator, indicating an attempt to violate a CPCON 3 directive. Upon isolating the affected host, a verification scan is conducted using Nmap, as shown in Fig. \ref{fig:WEB_Block_verify}, confirming that no unexpected ports are accessible.

Based on the correlation of multiple alert patterns, the orchestrator recommends escalating to CPCON level 2. Upon approval, the orchestrator executes an Ansible playbook to fully isolate the non-critical subnet, following a similar procedure to the prior HTTP/HTTPS traffic restriction.

A final Nmap scan confirms that no hosts within the non-critical subnet are reachable, with the exception of port 22 on the router, which remains accessible exclusively from the orchestrator’s IP address to support administrative access and potential restoration procedures. The policy repository is subsequently updated to indicate that CPCON~2 enforcement has been \textit{verified}, seen in the bottom section of Fig.~\ref{fig:Final_state}.

\section{Discussion}\label{Discussion}

We recognize that the orchestration prototype is a modest first step toward standardizing and automating CPCON operation at the platform level. In the following, we discuss several essential extensions that are required for a fleet-wide deployable system. 

\subsection{Verifying and Reporting of Compliance}
To support command and control across echelons, the orchestrator must reliably report policy compliance and operational status. Standardizing the monitoring infrastructure is critical to building a coherent operational picture. This includes automating the collection, aggregation, and reporting of policy enforcement data via standardized modules deployed across the DoDIN. These modules should feed relevant data to higher echelons, enabling systematic verification of policy adherence and improving situational awareness. Designing the monitoring components to operate across diverse system configurations supports seamless integration into a Common Operational Picture (COP).

\subsection{Scaling to DoDIN-wide Operations}
This work demonstrates the feasibility of implementing CPCON principles within a small network environment. Scaling these principles across the entire DoDIN could significantly improve its ability to defend against advanced cyber threats. Given the vast number of connected hosts on the DoDIN, applying orchestration techniques at scale presents substantial challenges. However, prior work has shown that orchestration latency increases linearly with the number of hosts~\cite{lukaszewski2023software}, suggesting that such an approach remains viable even in large-scale environments. A platform-agnostic design will be essential to ensure interoperability across heterogeneous systems.

\subsection{Supporting External Updates}
Effective orchestration requires the capability to ingest external updates, such as threat intelligence feeds and evolving policy guidelines. A key example includes Computer Tasking Orders (CTOs) issued by JFHQ-DoDIN, which convey authoritative directives and updated threat intelligence to DoD components. These updates must be integrated into the policy repository and disseminated to orchestrator modules in near real-time. Incorporating CTOs enables the system to dynamically adapt to emerging threats and ensures that policy enforcement remains aligned with current operational priorities and command-level guidance.


\subsection{Training}
Utilization of the orchestrator will reside with the local network commander, preserving tactical flexibility and preventing conflict with ongoing operations. Integration should be validated through fleet training events like Information Warfare Advanced Tactical Training and Composite Training Unit Exercise. which provide varied operational conditions to refine TTPs and reinforce C2 between the CWC and ship’s force. These exercises also test system interoperability within a system-of-systems framework and support improved network security practices among ITs.

\section{Conclusion}\label{Conclusion}
This paper presented a prototype CPCON orchestrator that automates the deployment and verification of policy-driven security actions across multiple hosts. Using an automated engine and a centralized policy repository, the system was shown to receive alerts, deploy response modules, recommend CPCON level escalation, and verify compliance in a virtual network testbed. Results suggest that the approach scales linearly with host count, supporting its feasibility in larger environments such as the DoDIN. Future work will focus on scaling to operational environments, standardizing monitoring, and integrating external threat intelligence to enhance orchestration effectiveness.


\end{document}